\documentclass[aps,prb,preprint,superscriptaddress,showpacs,amsmath]{revtex4}
\usepackage{graphicx,epsfig}

\begin{document}
\title{Electrical control of a laterally ordered InAs/InP quantum dash array}
\author{B. Al\'en}
\email[]{benito.alen@imm.cnm.csic.es}

\author{D. Fuster}

\author{I. Fern\'andez-Mart\'\i{}nez}

\affiliation{IMM, Instituto de Microelectr\'onica de Madrid (CNM, CSIC), Isaac Newton 8,
 28760 Tres Cantos, Madrid, Spain.}

\author{J. Mart\'\i{}nez-Pastor}

\affiliation{ICMUV, Instituto de Ciencia de Materiales, Universidad de Valencia,
  P.O. Box 2085, 46071 Valencia, Spain.}

\author{Y. Gonz\'alez}

\author{F. Briones}

\author{L. Gonz\'alez}

\affiliation{IMM, Instituto de Microelectr\'onica de Madrid (CNM, CSIC), Isaac Newton 8,
 28760 Tres Cantos, Madrid, Spain.}

\date{\today}

\begin{abstract}
We have fabricated an array of closely spaced quantum dashes starting from a planar array of self-assembled semiconductor quantum wires.
The array is embedded in a metallic nanogap which we investigate by micro-photoluminescence as a function of a lateral electric field. We
demonstrate that the net electric charge and emission energy of individual quantum dashes can be modified externally with performance
limited by the size inhomogeneity of the self-assembling process.
\end{abstract}

\pacs{78.67.Lt, 78.67.Hc, 73.63.Nm, 73.63.Kv}

\maketitle

A single semiconductor quantum dot (QD) is often considered the solid state analog of a single atom or ion since, in both, the electrons
occupy discrete energy levels which can be addressed individually using suitable optical fields. As a consequence, several solid state
quantum computation schemes have been proposed which exploit the spin and/or charge degrees of freedom of a single dot, using the
biexciton-exciton ladder,~\cite{Troiani2000,Li2003} or of several dots coupled by the exchange
interaction,~\cite{Piermarocchi2002,Petta2005} the light-matter interaction,~\cite{Imamoglu1999} the dipolar F\"{o}rster
interaction,~\cite{Hettich2002,Lovett2003} or the coherent tunnelling.~\cite{Robledo2008,Kim2008}

In most cases, the realization of an scalable multiparticle entangled state (a quantum byte) relies on the fabrication of an array of
closely spaced QD with good optical and electrical control over the single spin and/or charge states. The elementary block, the quantum dot
molecule (QDM), has been already demonstrated arranging vertically~\cite{Krenner2005,Stinaff2006} or laterally~\cite{Beirne2006} two
self-assembled QD. However, the fabrication of large, optically active, quantum dot arrays still remains a challenging task. In the last
years, quantum dot arrays or chains have been fabricated combining self-assembling methods with different patterning
processes,~\cite{Lee2001,Sun2002,Lefebvre2002,Alonso-Gonzalez2007,Sugaya2007} strain engineering of the
substrate,~\cite{Mano2002,Mazur2003} inclusion of misfit dislocations,~\cite{Leon2002,Welsch2008} or cleaved edge overgrowth
processes~\cite{Bauer2004}. The optical quality of these QDs has been assessed by characterization of their ensemble photoluminescence, and
increasing efforts are being directed to investigate the properties of individual QDs in the array~\cite{Sugaya2007,Bauer2004} or to
control them by applying a lateral electric field.~\cite{Reimer2008a}

In the following, we present a new method to fabricate an array of elongated quantum dots or quantum dashes (QDh) based on the post-growth
processing of a sample containing a single layer of self-assembled quantum wires (QWRs). The QDh array was embedded between lateral
electrodes and the emission properties of the individual QDh were investigated by micro-photoluminescence ($\mu$PL) as a function of the
lateral electric field. Our method is based on standard top-down semiconductor wafer processing of the as-grown substrate and therefore it
is potentially scalable. Yet, our study will also reveal the limitations imposed by current self-assembling processes to fabricate large
arrays of coupled quantum dots.

\begin{figure}[htb]
\begin{center}
\includegraphics[width=46 mm]{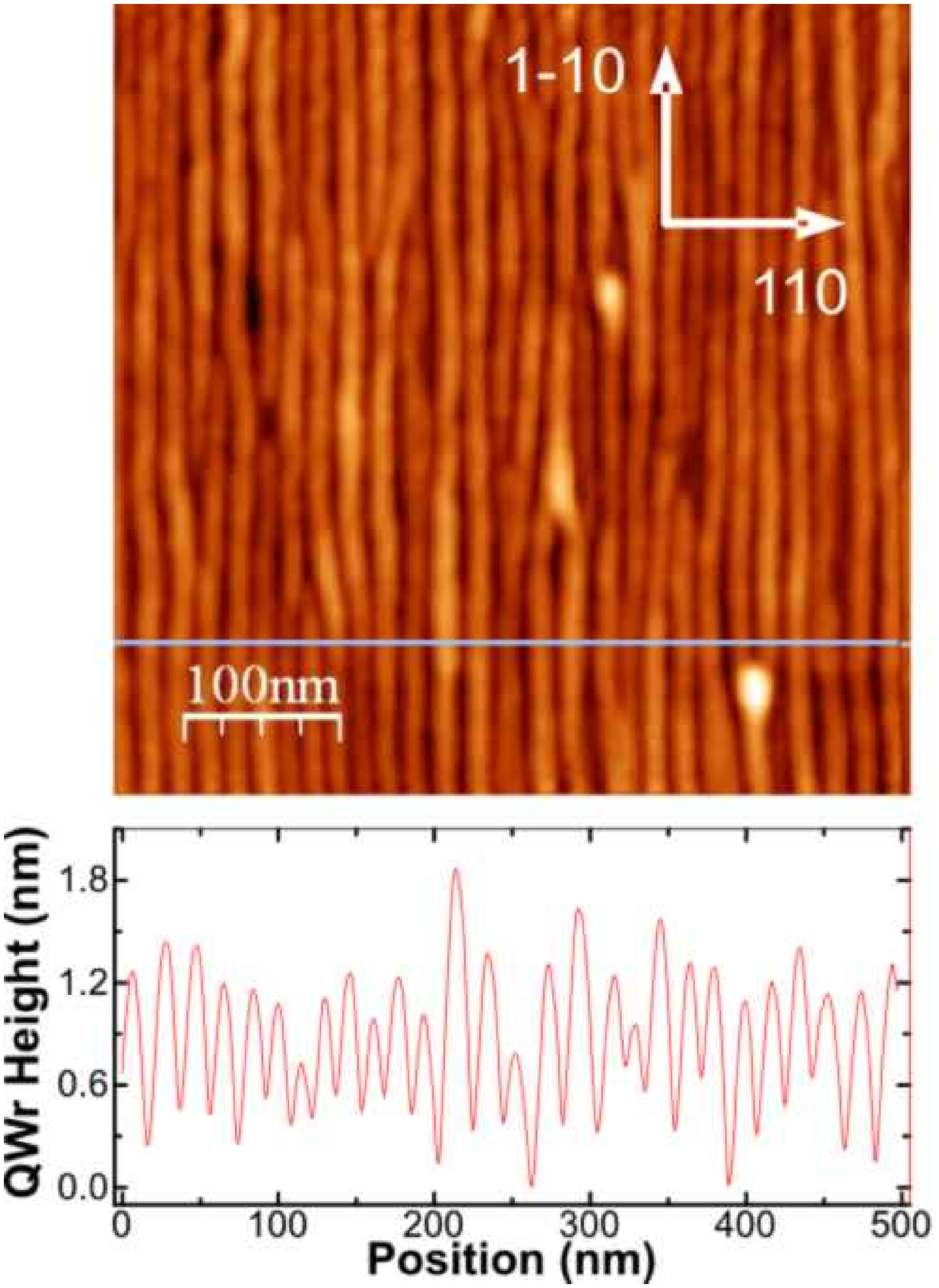}%
\caption{B. Al\'en et al.} \label{Fig1}
\end{center}
\end{figure}

The InAs self-assembled QWRs were fabricated by solid source molecular beam epitaxy (MBE) on InP (001). More details about the QWR growth
procedure can be found elsewhere.~\cite{Fuster2005} For our present purpose, their most relevant physical properties are collected in
Figure~\ref{Fig1}. The Atomic Force Microscopy (AFM) characterization of the uncapped sample reveals $1.2\pm0.3$ nm-thick self-assembled
QWRs disposed along the (1-10) crystal direction and forming a continuous array across their perpendicular direction. The pitch period of
the array (peak to peak) is $18\pm2$ nm and the average inter-QWr spacing at half-maximum is just $\sim4$ nm. The uniformity of the array
over large distances and the small separation are desirable characteristics for the fabrication of a QDh array and motivate this work.

\begin{figure}[htb]
\begin{center}
\includegraphics[width=80 mm]{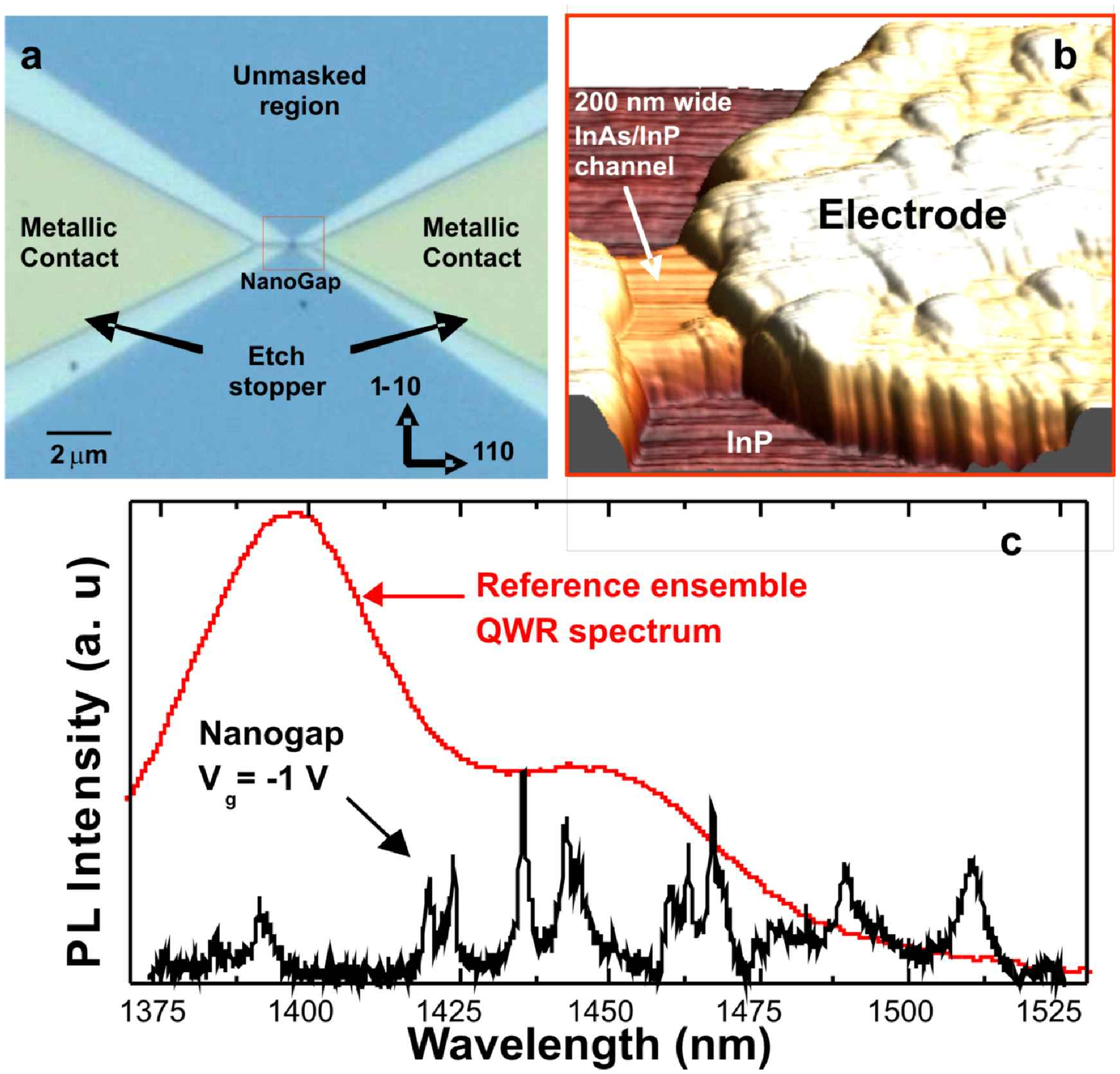}%
\caption{B. Al\'en et al.} \label{Fig2}
\end{center}
\end{figure}

To fabricate the array, a sample was grown with the QWRs covered by 20 nm of InP and, next, two gate electrodes (10 nm Cr + 25 nm Au) were
defined over the InP surface using electron beam lithography (EBL) and lift-off methods. The separation between the two electrodes along
the (110) crystal direction was set to $\approx$200 nm meaning that $\approx$10 QWRs were embedded in the nanogap at this stage. The device
geometry was delimited further defining etch stoppers in a second EBL step as shown in Figure~\ref{Fig2}(a). Both, the InP capping layer
and the QWR layer were dry etched in the unmasked regions by reactive ion beam etching of InP using CH$_4$:H$_2$ buffered with N$_2$. After
the processing, the final layout of the device consists of a narrow rectangular channel (200$\times$300 nm) as shown in the AFM micrograph
of Figure~\ref{Fig2}(b) where the QDh are allocated parallel to the electrode flat edges.

The $\mu$PL spectrum of the nanodevice was analyzed using a fiber based confocal microscope working at 6 K. Continuous-wave excitation
below the InP band gap was provided by a Ti-Sapphire laser ($\lambda_{exc}=980$ nm) coupled to the excitation fiber. The
micro-photoluminescence was collected with a different fiber and detected using a multichannel InGaAs photodiode array attached to a 0.5 m
focal length spectrograph.

\begin{figure}[htb]
\begin{center}
\includegraphics[width=90 mm]{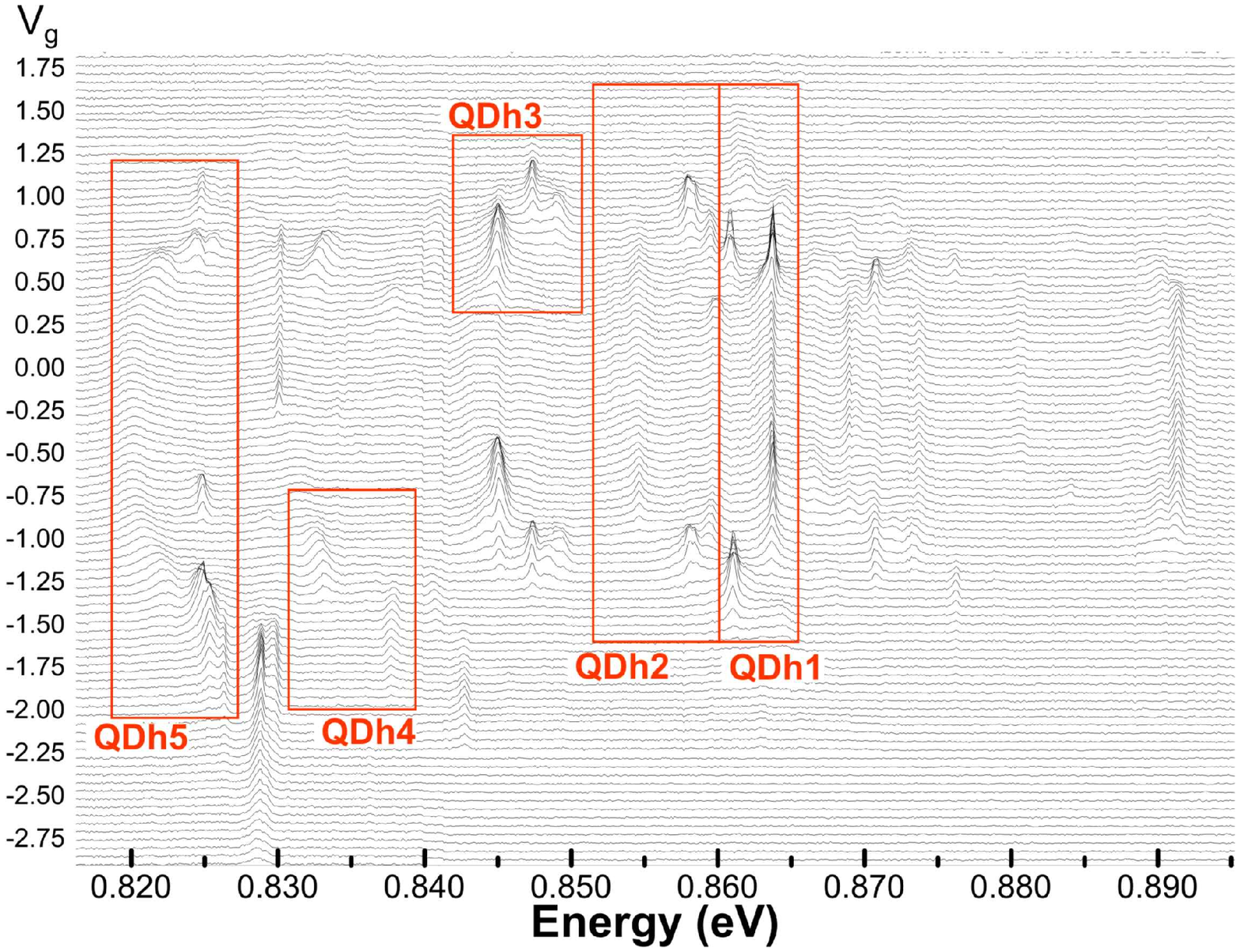}%
\caption{B. Al\'en et al.} \label{Fig3}
\end{center}
\end{figure}

Figure~\ref{Fig2}(c) shows a $\mu$PL spectrum obtained between the electrodes together with a reference ensemble PL spectrum recorded far
away in a non-etched region. The latter is characteristic of the multimodal distribution of QWR heights observed by AFM.~\cite{Alen2001}
Meanwhile, the sharp spectral features recorded in the nanogap region must be associated to the emission of individual nanostructures
allocated inside.~\cite{Alen2006} In this situation, a lateral electric field can be applied to investigate the bias evolution of the
observed resonances, as shown in Figure~\ref{Fig3}. Since the gate electrodes are arranged symmetrically on the surface, we observe that
the dependence is approximately symmetrical with respect to $V_g=0$ V.

Generally, although the optical excitation generates matched electron-hole pairs,  the electron and hole occupation in a given QDh is
unequal and depends on the impurity background and on the different probabilities for electron and hole capture.~\cite{Munoz-matutano2008}
In our case, Hall resistance measurements reveal a n-type character ($N_D\approx1\times10^{16}$ cm$^{-2}$) for the InP barriers favoring
the formation of negatively charged excitons when no bias is applied.~\cite{Alen2008} Also, carriers photogenerated at high energies must
drift in the illuminated area and can be trapped in shallow traps associated to thickness fluctuations of the wetting layer before being
captured in the QDh. Since the trapping efficiency is larger for holes, this mechanism can also lead to unbalanced charge configurations in
the QDh.~\cite{Moskalenko2006} In both situations, the number of confined carriers can be controlled applying a lateral electric field. The
role of the bias is, for moderate values, to release the trapped charge which can now be collected in the QDh restoring the neutrality of
the capture process~\cite{Moskalenko2006} and, for larger positive or negative voltages, to bend the conduction and valence bands and
enable the tunnelling of confined carriers out of the QDh.~\cite{Alen2007}

\begin{figure}[htb]
\begin{center}
\includegraphics[width=80 mm]{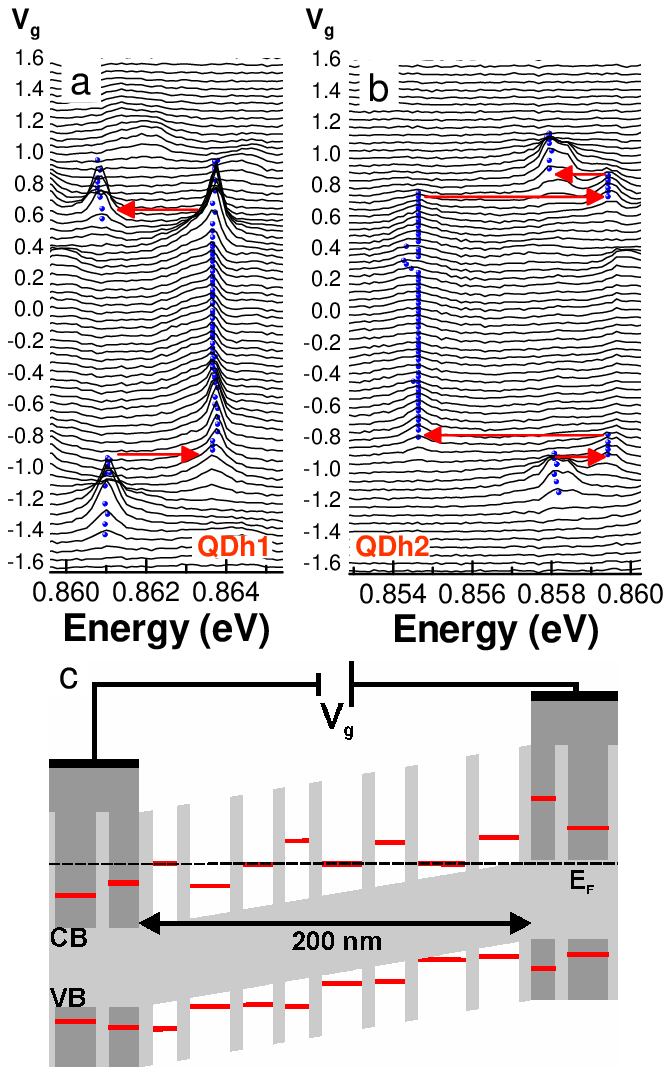}%
\caption{B. Al\'en et al.} \label{Fig4}
\end{center}
\end{figure}

The behavior just described explains the bias evolution shown in Figure~\ref{Fig4}(a) where we have magnified a narrow region of the
spectrum containing the emission of a single QDh (highlighted as QDh1 in Figure~\ref{Fig3}). A single resonance centered at 0.8636 eV
characterizes the spectrum for $-0.7<V_g<0.7$ V. In the midst of this range the intensity is smaller and increases towards both limits
where it finds a maximum thanks to the enhanced carrier drift velocity.~\cite{Moskalenko2006} Further positive or negative bias results
indistinctly in a rapid quench of this resonance and the appearance of a new peak centered at 0.8608 eV. This happens after a change of the
QDh net charge which shifts the ground state emission to lower energies due to the new balance between attractive and repulsive Coulomb
terms.~\cite{Alen2005} Given the abrupt character of the transition and the n-type character of the sample, the charging mechanism must be
related to a tunnelling event occurring at this voltage as the fermi level crosses the conduction band levels. The positions of the energy
jumps can be used thereafter to identify which resonances correspond to different QDh.~\cite{Alen2005} If we keep increasing $|V_g|$, as
shown in Figure~\ref{Fig3}, all the emission lines finally disappear due to the fast tunnelling of carriers out of the QDh before they can
recombine radiatively.~\cite{Alen2007}

The general picture described for QDh1 is valid to explain the evolution of other QDhs in the array. For instance, Figure~\ref{Fig4}(b)
shows the evolution of QDh2 which, in the same voltage range, suffers two charging events accompanied by two energy shifts of opposite
sign. In our present device geometry, the charge state of the QDh depends mostly on the alignment of its confined levels with the chemical
potential and therefore on the particular size and position of the QDh within the array [Fig.~\ref{Fig4}(c)].~\cite{Reimer2008} Clearly,
although the net electric charge can be varied in different QDh, its precise value can not be set in each one independently.  This
limitation, imposed by the size inhomogeneity inherent in self-assembled growth methods, could be solved in the future using individual
gating technologies.

The average separation of 18 nm between QDh centers and the small $\approx4$ nm edge to edge distance should be adequate to observe quantum
tunnelling phenomena in this system.~\cite{Krenner2005,Stinaff2006,Beirne2006} Yet, the mere observation of charging events does not imply
a charge exchange between the QDh lined up between the electrodes. As stated above, the bias dependency of the carrier transport and
tunnelling rates can explain our results as well. In the following, we will focus on the anomalous Stark shift found in different regions
of the QDh array emission spectrum and discuss whether it can be interpreted in terms of indirect transitions among coupled QDh.

\begin{figure}[htb]
\begin{center}
\includegraphics[width=80 mm]{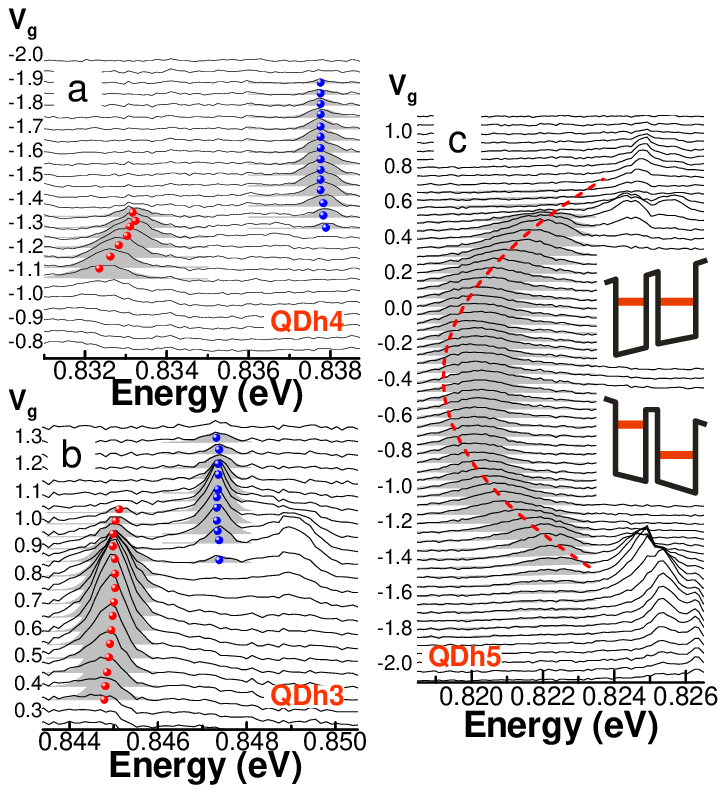}%
\caption{B. Al\'en et al.} \label{Fig5}
\end{center}
\end{figure}

Figure~\ref{Fig5}(a) shows two resonances corresponding to QDh4 in its negative voltage range. Increasing the bias from $-1.0$ V, the
resonance at the left shifts to higher energies finding a maximum at $-1.3$ V. The remarkable blue-shift cannot be explained from the
quantum confined Stark effect of a direct exciton and distinguishes this resonance from those shown in figure~\ref{Fig4}(a) and (b). At
$-1.4$ V, the peak disappears and a new resonance becomes visible $\sim 4.7$ meV apart towards higher energies. The rest of
Figure~\ref{Fig5} depicts the dependence found for QDh3 in its positive bias range showing a similar pair of resonances split by $\sim 2.4$
meV [Fig.~\ref{Fig5}(b)], and the full evolution of QDh5 which demonstrates that the emission blue-shift is indeed symmetric around
$V_g\sim0$ V [Fig.~\ref{Fig5}(c)].

Large energy splittings and anomalous Stark shifts, including blue-shifts similar to those shown in Figures~\ref{Fig5}(a) and (b), are
spectral characteristics of vertically coupled QD molecules when an electron becomes delocalized between two QDs giving rise to resonance
anticrossings.~\cite{Krenner2005,Krenner2006} For lateral molecules, however, the wavefunction overlap is smaller, quantum tunnelling
splittings greater than a few hundred $\mu$eV are unlikely, and anticrossings are hardly visible.~\cite{Beirne2006} In addition, we observe
the same evolution for opposite polarities of the bias [Fig.~\ref{Fig3}]. This is difficult to match within a resonant tunnelling scheme
since two given electronic levels which are mutually aligned at $V_0$, must be necessarily separated at $-V_0$, as shown schematically in
Figure~\ref{Fig5}(c) for an asymmetric QDM. The possibility of two different anticrossings involving neutral and charged excitons as
observed in vertical QDM,~\cite{Stinaff2006,Krenner2006} can also be ruled out in our case where we find the same lines and splittings for
both bias. We conclude that the anomalous Stark shift can not be related to an indirect transition. Recently, blue-shifting resonances have
been observed in single InAs/InP QDs embedded between lateral electrodes.~\cite{Reimer2008a}  A theoretical model was able to explain them
due to the field induced screening of the electron-hole interaction in a confined biexciton.~\cite{Korkusinski2009} Alternately, the
observed behavior could be related to a charging effect. As the device approaches the flat band condition at zero bias, more electrons
populate the conduction band levels. In QWRs, such scenario of many electrons and few holes leads to the renormalization of the band gap
energy and shifts the emission in the direction observed here.~\cite{Akiyama2002} In QDs, highly charged exciton peaks also exhibit
anomalous shifts in the flat band vicinity.~\cite{Alen2005} Both interpretations, involving biexciton or highly charged exciton
contributions, are consistent with the symmetric bias dependence. The latter also agrees with the n-type character of the InP barriers and
the low excitation power used in our experiments.

Our results clearly show the limiting aspects of self-assembled growth methods for the fabrication of large arrays of coupled QDs. In our
case, a possible spatial mismatch due to carrier localization along the QDh long axis has to be added to the electronic levels misalignment
which often hampers the observation of quantum tunnelling phenomena. The recent development of pre-patterning methods where the position of
QDs can be controlled without affecting their optical properties should encourage further work in this
field.~\cite{Reimer2008a,Martin-Sanchez2009b}

In conclusion, we have fabricated an array of closely spaced quantum dashes by post-growth processing of a single layer of InAs/InP
self-assembled quantum wires grown by solid source molecular beam epitaxy. The array was embedded in a lateral field effect structure which
allowed the study of the emission spectrum of individual quantum dashes as a function of the applied bias. We have demonstrated that the
net electric charge can be controlled and depleted in the individual nanostructures and that the energy position of their electronic levels
can be modified. Our study reveals that size fluctuations prevent however an accurate control of these parameters necessary to fabricate
large arrays of coupled QDs using this technology.

The authors acknowledge financial support by Spanish MEC and CAM through grants NANINPHO-QD (TEC2008-06756-C03-01/03), NANOCOMIC
(S-505/ESP/000200) and Consolider-Ingenio 2010 QOIT (CSD2006-0019).


\newpage
\Large{\textbf{List of figures}}\\

\vspace{6 mm} \small

\textbf{Figure 1.-} Top panel: 500$\times$500 nm AFM image of an uncapped sample showing the initial QWR planar array. Bottom panel:
Profile along the (110) crystal direction of the same AFM image.

\textbf{Figure 2.-} (a) Optical image of the device before the dry etching step. (b) AFM micrograph of the fully processed 200 nm wide
nanogap containing the array of InAs/InP QDh. (c) Ensemble PL of the unprocessed sample and $\mu$PL spectrum in the nanogap obtained at 6
K.

\textbf{Figure 3.-} Full evolution of the $\mu$PL spectrum recorded in the nanogap region with external bias. Highlighted regions enclose
the emission of individual QDhs as discussed in the text.

\textbf{Figure 4.-} a) and b) Different spectral regions where the coulomb blockade effects dominate the bias evolution of the $\mu$PL
spectrum. The horizontal arrows indicate the voltages where the net electric charge changes in each QDh. c) Schematics of the device band
structure along the 110 crystal direction in the QDh array region.

\textbf{Figure 5.-} a), b) and c) Different spectral regions where individual resonances experience a blue-shift with increasing bias
(solid dots in a) and b) indicate the positions of the maxima performing gaussian fits, dashed line in c) is a guide to the eye). The
evolution is symmetric around $V_g=0$ V as observed in panel c) and Fig.~\ref{Fig3}.

\newpage
\newpage
\newpage
\newpage
\newpage

\end{document}